\documentclass[prb,a4paper,twocolumn,superscriptaddress,amsmath,showpacs,showkeys]{revtex4}
\pdfoutput=1

\usepackage{graphicx}
\usepackage{dcolumn}
\usepackage{textcomp}  

\newcommand{\be}{\begin{equation}}
\newcommand{\ee}{\end{equation}}
\newcommand{\aref}[1]{App.~\ref{#1}}
\newcommand{\eref}[1]{Eq.~(\ref{#1})}
\newcommand{\Eref}[1]{Equation (\ref{#1})}
\newcommand{\fref}[1]{Fig.~\ref{#1}}
\newcommand{\Fref}[1]{Figure \ref{#1}}
\newcommand{\rref}[1]{Ref.~\onlinecite{#1}}
\newcommand{\sref}[1]{Sec.~\ref{#1}}
\newcommand{\tref}[1]{Tab.~\ref{#1}}
\newcommand{\sub}[1]{_\mathrm{#1}}
\newcommand{\rmd}{\mathrm{d}}
\newcommand{\rme}{\mathrm{e}}

\DeclareMathOperator{\erf}{erf}

\begin{document}

\title{Measurement of work in single-molecule pulling experiments}
\author{Alessandro Mossa}
\email{mossa@ub.edu}
\affiliation{Departament de F\'{\i}sica Fonamental, Facultat de F\'{\i}sica, Universitat de Barcelona,\\ Avinguda Diagonal 647, 08028 Barcelona, Espa\~na}
\author{Sara de Lorenzo}
\affiliation{Departament de F\'{\i}sica Fonamental, Facultat de F\'{\i}sica, Universitat de Barcelona,\\ Avinguda Diagonal 647, 08028 Barcelona, Espa\~na}
\affiliation{CIBER de Bioingenier\'{\i}a, Biomateriales y Nanomedicina, Instituto de Salud Carlos III, Madrid, Espa\~na}
\author{Josep Maria Huguet}
\affiliation{Departament de F\'{\i}sica Fonamental, Facultat de F\'{\i}sica, Universitat de Barcelona,\\ Avinguda Diagonal 647, 08028 Barcelona, Espa\~na}
\author{Felix Ritort}
\email{ritort@ffn.ub.es}
\affiliation{Departament de F\'{\i}sica Fonamental, Facultat de F\'{\i}sica, Universitat de Barcelona,\\ Avinguda Diagonal 647, 08028 Barcelona, Espa\~na}
\affiliation{CIBER de Bioingenier\'{\i}a, Biomateriales y Nanomedicina, Instituto de Salud Carlos III, Madrid, Espa\~na}

\begin{abstract}
A main goal of single-molecule experiments is to evaluate equilibrium
free energy differences by applying fluctuation relations to
  repeated work measurements along irreversible processes. We quantify the
  error that is made in a free energy estimate by means of the Jarzynski
  equality when the accumulated work expended on the whole system
  (including the instrument) is erroneously replaced by the work
  transferred to the subsystem consisting of the sole molecular
  construct. We find that the error may be as large as 100\%, depending
  on the number of experiments and on the bandwidth of the data
  acquisition apparatus. Our theoretical estimate is validated by
  numerical simulations and pulling experiments on DNA hairpins using 
  optical tweezers.
\end{abstract}


\maketitle

\section{Introduction} \label{sec:intro}

In a typical single-molecule pulling ex\-per\-i\-ment\cite{Ritort:2006aa}, an individual molecular construct is stretched by means of a device (e.g., optical or magnetic tweezers, atomic force microscope (AFM), etc.) able to measure both the applied force, usually on the piconewton scale, and the end-to-end molecular extension, typically expressed in nanometers. Many interesting kinetic and thermodynamical properties\cite{Hyeon:2007aa,Manosas:2005aa,Manosas:2006aa} of the stretching process can be inferred from the resulting force-extension curve (henceforth, FEC); in particular, the free energy difference between the folded and the unfolded state can be evaluated by exploiting a well-known result of nonequilibrium thermodynamics, the Jarzynski equality\cite{Jarzynski:1997aa}:
\be \label{eq:JE}
	\beta W\sub{rev}=-\log\langle\exp[-\beta W(\Gamma)]\rangle_\Gamma \,,
\ee
where $W(\Gamma)$ is the amount of work performed on the system throughout the stretching process $\Gamma$, $\beta$ is as usual the inverse of the thermal energy $k\sub{B}T$, and $W\sub{rev}$ is the reversible work, i.e., the work needed to perform the pulling experiment in quasi-equilibrium conditions. Since a single molecule is a small system\cite{Ritort:2008aa,Marini-Bettolo-Marconi:2008aa,Ritort:2007aa}, $W(\Gamma)$ is affected by thermal fluctuations; the angular brackets $\langle\cdots\rangle_\Gamma$ thus stand for an average over all possible realizations of the same experimental protocol. In fact, a generalization of the Jarzynski equality due to Hummer and Szabo\cite{Hummer:2001aa,Hummer:2005aa} makes it possible to reconstruct 
the whole free energy landscape as a function of the molecular extension\cite{Braun:2004aa,Imparato:2006aa,Hyeon:2008aa}. This program has been successfully applied to the experimental study of multi-domain proteins\cite{Harris:2007aa,Imparato:2008aa}.

Many a research has been devoted to the practical difficulties that arise when \eref{eq:JE} is applied to the free energy reconstruction problem, e.g., the bias induced by the finite number of experimental attempts\cite{Gore:2003aa}, the role played by the resolution of the measuring apparatus\cite{Rahav:2007aa}, or the effect of instrument noise and experimental errors\cite{Maragakis:2008ab}. The present article deals with yet another possible source of error, which, though already known, has generally been dismissed as negligible without a compelling argument. The point is that in most experimental settings the molecular extension is \emph{not} the proper control parameter, so that it is not correct to interpret the area below the FEC as the work that appears in \eref{eq:JE}\cite{Schurr:2003aa}. If the control parameter is the total distance the area under the force-distance curve (FDC) should be used instead. 

Here we thoroughly analyze under which conditions the use of the wrong
definition for the work can appreciably affect the estimate of free
energy differences by means of \eref{eq:JE}. The conclusion, in a
nutshell, is that the error induced by the substitution may be as large
as 100\%, depending on the number of experiments and on the data
acquisition frequency. Also important are the details of the data
analysis procedure: how the integration extrema are chosen, what
method is used to integrate the FEC and how different FECs are aligned  to correct for instrumental drift effects.

The paper is organized as follows: First, we get some theoretical insight by considering our problem in its simplest possible setting (\sref{sec:toy}). Then, we validate our conclusions with an experimental test   
implemented with optical tweezers and DNA hairpins (\sref{sec:exp}). A recapitulation of our results (\sref{sec:end}) and an appendix with some technicalities round off this article. 

\section{A toy model} \label{sec:toy}

A detailed model for single-molecule experiments with optical tweezers has been discussed elsewhere\cite{Manosas:2005aa}. Here we consider a simplified version of it, that conserves only the physical features directly relevant to our problem. Although the toy model in this section is phrased in the optical tweezers language, it takes no effort to translate it into an AFM nomenclature, the mathematics being just the same.

\begin{figure}
   \includegraphics[width=8.6cm]{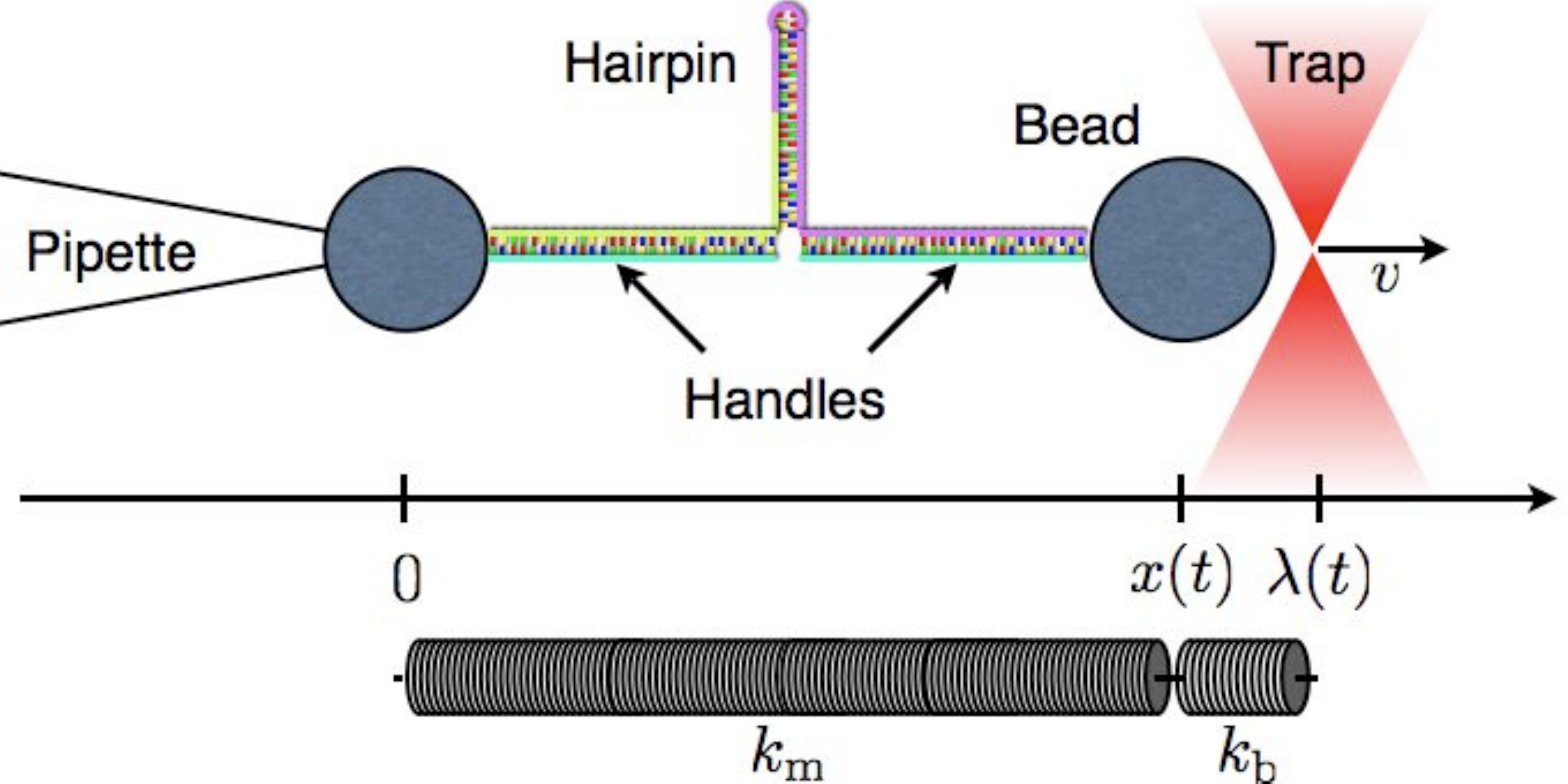}
    \caption{Schematic definition of the model under study. The pipette is at rest with respect to the thermal bath, while the trap is moving with velocity $v$. The trap and the system molecule + handles are approximated by two harmonic potentials with stiffness $k\sub{b}$ and $k\sub{m}$, respectively. The rest length of the trap spring $k\sub{b}$ is zero, while the rest length of the molecule spring $k\sub{m}$ is $\ell_0$ if the hairpin is closed ($\varsigma=0$) and $\ell_1$ if it is open ($\varsigma=1$).} 
   \label{fig:setup} 
\end{figure}

In our model, graphically depicted in \fref{fig:setup}, the optical trap
is moved by the experimenter, hence the proper control parameter is the
trap--pipette distance $\lambda$, while the end-to-end molecular
extension is a quantity subject to fluctuations denoted by $x$. The trap
is an harmonic potential with stiffness $k\sub{b}$, while $k\sub{m}$ is
the stiffness of the molecular construct comprising hairpin and
handles. Given a fixed value of the control parameter $\lambda$, the
state of the system is specified by the pair $(x,\varsigma)$, where
$\varsigma$ is a label taking values 0 if the hairpin is closed (or
  folded) and 1 if it is open (or unfolded).
The hairpin itself is a pure two-state system\cite{Ritort:2002aa} whose
state-dependent length is $\ell_\varsigma$. The bead is thus subject to
the net force
\be \label{eq:force}
	f\sub{t}(x,\varsigma)=k\sub{b}(\lambda-x)-k\sub{m}(x-\ell_\varsigma) \,.
\ee
It is convenient to introduce the total stiffness $k\sub{t}\equiv k\sub{b}+k\sub{m}$ and the equilibrium position (defined by the condition $f\sub{t}(x\sub{eq},\varsigma)=0$)
\be
	x\sub{eq}(\varsigma)=\frac{k\sub{b}\lambda+k\sub{m}\ell_\varsigma}{k\sub{t}} \,,
\ee
so that \eref{eq:force} can be rewritten as
\be
	f\sub{t}(x,\varsigma)=-k\sub{t}[x-x\sub{eq}(\varsigma)] \,.
\ee
The relaxation time of the velocity autocorrelation
function $\tau=m/\gamma$ ($m$ being the mass and $\gamma$ the friction coefficient of the bead in the trap) is small enough compared to the duration of the experiment that we can assume \emph{mechanical equilibrium}\cite{Astumian:2007aa}, i.e.\ the average value of the total force $\langle f\sub{t}(t)\rangle$ is zero.
The Hamiltonian function is given by 
\be \label{eq:Ham}
	H^{(\lambda)}(x,\varsigma)={\textstyle \frac{1}{2}}k\sub{b}(\lambda-x)^2+{\textstyle \frac{1}{2}}k\sub{m}(x-\ell_\varsigma)^2+\varsigma\Delta G_0 \,,
\ee
where $\Delta G_0$ is the free energy difference between the open and closed states of the hairpin in the absence of applied force.
The analytic solution to the equilibrium thermodynamics of this model is summarized in \aref{sec:therm}.

The transitions of the hairpin are governed by a simplified Kramers--Bell kinetics\cite{Tinoco:2004aa}, with rates for opening $k_\to$ or closing $k_\gets$ given by 
\begin{subequations}
\begin{align}
	k_\to&=k_0\exp\left(\frac{w_0f_0(x)}{k\sub{B}T}\right) \,, \\
	k_\gets&=k_0\exp\left(\frac{-w_1f_1(x)+\Delta G_0}{k\sub{B}T}\right) \,,
\end{align}
\end{subequations}
where $w_0$ and $w_1$ represent the distances from the barrier to the closed and the open states, respectively, $f_0$ and $f_1$ are two functions of $x$ with physical dimensions of a force, and $k_0$ is the attempt frequency. The rates just defined must respect the detailed balance condition
\be
	\frac{k_\to}{k_\gets}=\exp\left[-\frac{H^{(\lambda)}(x,1)-H^{(\lambda)}(x,0)}{k\sub{B}T}\right] \,,
\ee
for each $\lambda$ and for each $x$. This requirement implies
\be
	w_0f_0(x)+w_1f_1(x)=\frac{k\sub{m}}{2}(\ell_1-\ell_0)[2x-(\ell_1+\ell_0)] \,.
\ee
Our choice here is to take simply $f_0(x)=f_1(x)$, so that $w_0+w_1=\ell_1-\ell_0$.  

The dynamics of our model is ruled by the overdamped Langevin equation
\be \label{eq:LangEq}
	\gamma\frac{\rmd x}{\rmd t}=f\sub{t}(x(t),\varsigma)+\sqrt{2\gamma k\sub{B}T}\xi(t) \,,
\ee
where $\xi(t)$ is a Gaussian white noise
\begin{subequations}
\begin{align}
	\langle\xi(t)\rangle&=0 \,, \\
	\langle\xi(t)\xi(t')\rangle&=\delta(t-t') \,.
\end{align}	
\end{subequations}
The experimental protocol is defined by the choice of a function $\lambda(t)$. Here we consider a constant velocity pulling: $\lambda(t)=\lambda_0+vt$. 

\subsection{Accumulated vs.\ transferred work}

\begin{figure*}
   \includegraphics[width=8.6cm]{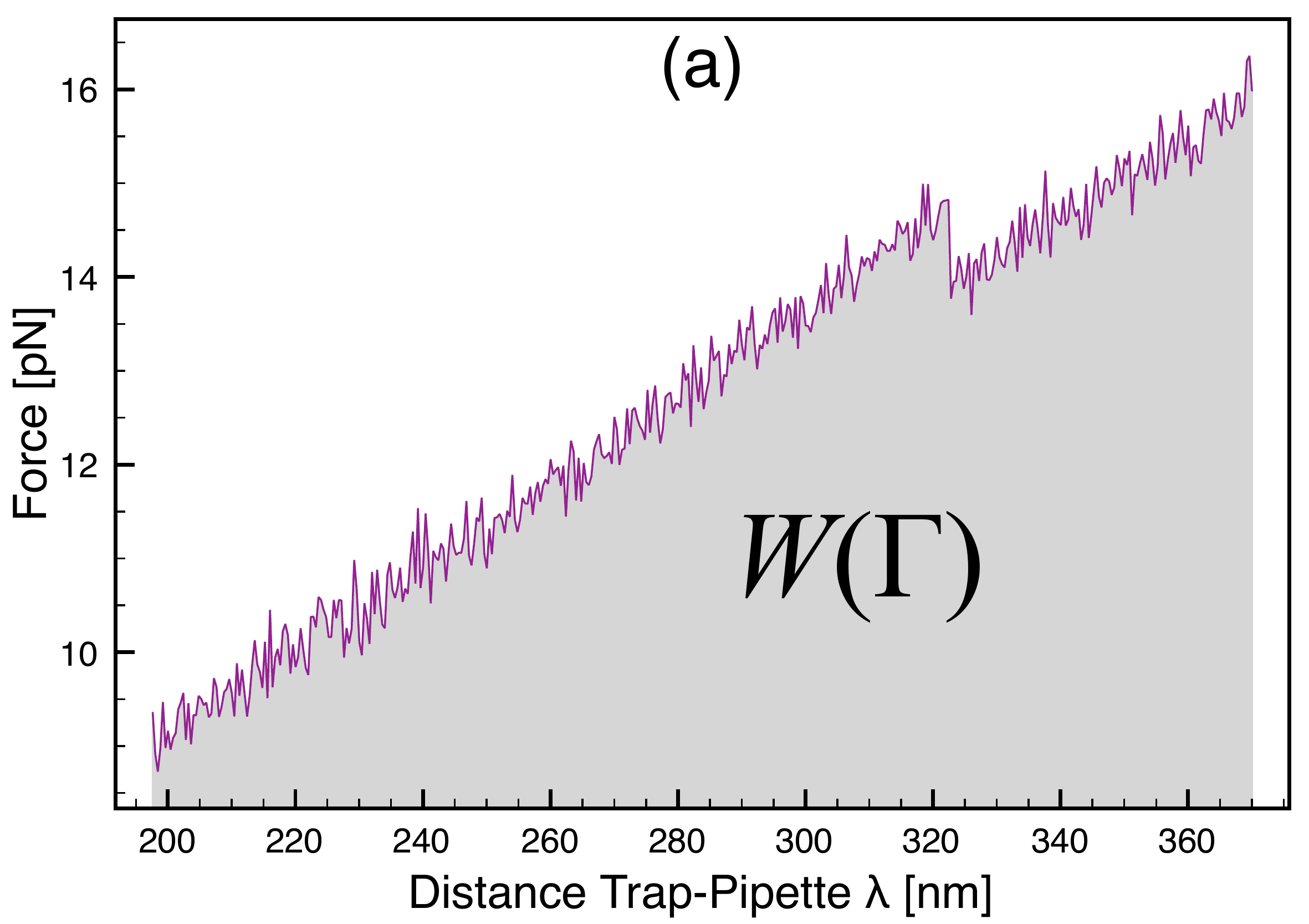} \hspace{0.3cm}
   \includegraphics[width=8.6cm]{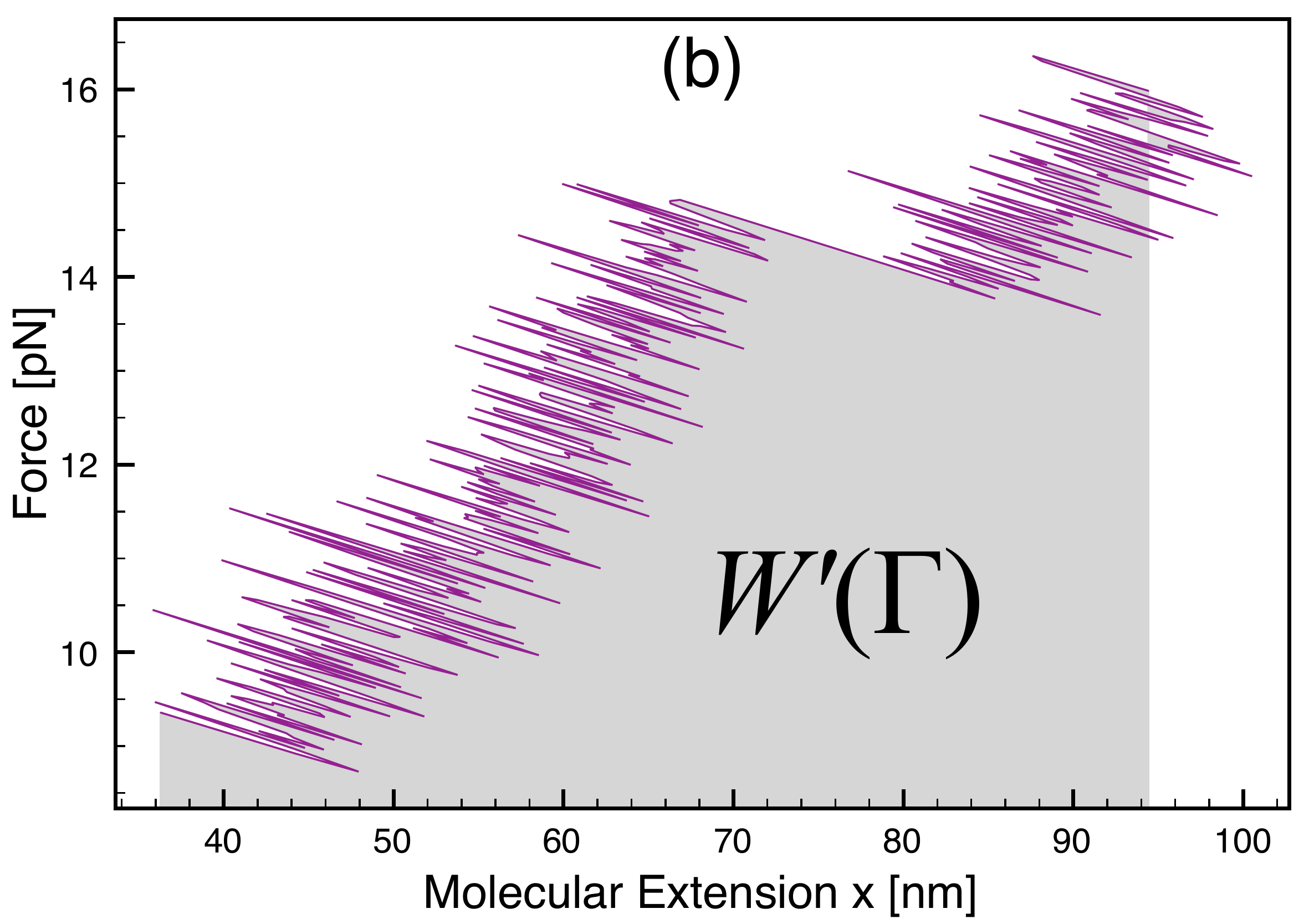}
    \caption{(a) A typical force-distance curve (FDC) obtained by numerical simulation of \eref{eq:LangEq}. The shaded area is equivalent to the accumulated work $W(\Gamma)$ [see \eref{eq:W(Gamma)}]. (b) The force-extension curve (FEC) associated to the pulling experiment represented in \fref{fig:FDCvsFEC}(a). The shaded area is equivalent to the transferred work $W'(\Gamma)$ [see \eref{eq:W'(Gamma)}].} 
   \label{fig:FDCvsFEC} 
\end{figure*}

For the toy model introduced in the previous section, $\lambda$ is the
control parameter, which can be directly manipulated, while the
molecular extension $x$ is subject to Brownian fluctuations. Therefore, the work performed on the system throughout a pulling experiment $\Gamma$ that starts at time $t\sub{i}$ from $\lambda=\lambda\sub{i}$ and terminates in $\lambda=\lambda\sub{f}$ at time $t\sub{f}=t\sub{i}+(\lambda\sub{f}-\lambda\sub{i})/v$ is properly defined as
\be \label{eq:W(Gamma)}
	W(\Gamma)\equiv\int_{\lambda\sub{i}}^{\lambda\sub{f}}\frac{\partial H^{(\lambda)}(x,\varsigma)}{\partial \lambda}\,\rmd \lambda=\int_{\lambda\sub{i}}^{\lambda\sub{f}}f\sub{b}(\lambda,x)\,\rmd \lambda \,.
\ee
where we used \eref{eq:Ham} and $f\sub{b}(\lambda,x)\equiv
k\sub{b}(\lambda-x)$ is the force induced by the displacement of the
bead in the trap. Such work is measured in practice as the area under
the force-distance curve [FDC, see \fref{fig:FDCvsFEC}(a)].  Note that for all single-molecule techniques that we are aware of, $f\sub{b}$ is actually the only one force experimentally measurable. In the following we will for simplicity drop the subscript and write $f$ instead of $f\sub{b}$.

The area under the FEC [see \fref{fig:FDCvsFEC}(b)], on the other hand, is what in \rref{Schurr:2003aa} is called transferred work [as opposed to the accumulated work $W(\Gamma)$]: 
\be \label{eq:W'(Gamma)}
	W'(\Gamma)\equiv\int_{x\sub{i}(\Gamma)}^{x\sub{f}(\Gamma)}f(\lambda,x)\,\rmd x \,,
\ee 
where $x\sub{i}$ and $x\sub{f}$ are the \emph{trajectory-dependent} values of the molecular extension at times $t\sub{i}$ and $t\sub{f}$, respectively.  

At each point along the trajectory $\Gamma$, the control parameter and the molecular extension are  related by
\be \label{eq:xvslambda}
	x = \lambda-\frac{f}{k\sub{b}} \,.
\ee
This implies the following relation between the area under a FDC and the area under the corresponding FEC:
\be \label{eq:W-W'}
	W(\Gamma)=W'(\Gamma)+\frac{f\sub{f}(\Gamma)^2-f\sub{i}(\Gamma)^2}{2k\sub{b}} \,,
\ee
where $f\sub{i}$ and $f\sub{f}$ are the (trajectory-dependent) initial and final values of the force, respectively. The difference between $W$ and $W'$ is therefore a pure \emph{boundary} term.

\subsection{The reversible work}

If we realize the pulling experiment in conditions of quasi-equilibrium, that is at infinitesimally small velocity $v\to0$, then we obtain the thermodynamic force-distance curve (TFDC), whose analytical expression is given by \eref{eq:TFDC}. The area under the TFDC is the reversible work $W\sub{rev}$, equal to the free energy difference between the final and initial states of the system. From an experimental perspective, however, the really interesting quantity is rather the free energy difference $\Delta G_0$ between the open and closed states of the hairpin at zero external force. According to \eref{eq:Wrev}, this is given by
\be \label{eq:DeltaE}
	\Delta G_0 = W\sub{rev}-\frac{\langle f\rangle\sub{f}^2-\langle f\rangle\sub{i}^2}{2k\sub{eff}} \,,
\ee
where $\langle f\rangle\sub{i(f)}$ is the equilibrium initial (final) value of the force, and $k\sub{eff}$ is the effective stiffness
\be
	\frac{1}{k\sub{eff}}=\frac{1}{k\sub{b}}+\frac{1}{k\sub{m}} \,.
\ee

The thermodynamic force-extension curve\cite{Manosas:2005aa} (TFEC) is the quasi-equilibrium pulling experiment plotted as a function of the molecular extension $x$. If we define $W'\sub{rev}$ as the area under the TFEC, then \eref{eq:W-W'} yields
\be \label{eq:DeltaE'}
	\Delta G_0 = W'\sub{rev}-\frac{\langle f\rangle\sub{f}^2-\langle f\rangle\sub{i}^2}{2k\sub{m}} \,.
\ee
So we see that either $W\sub{rev}$ or $W'\sub{rev}$ are equally useful
to extract the free energy of formation $\Delta G_0$ of the hairpin. The problem is that it is often unpractical (and sometimes impossible) to achieve quasi-equilibrium conditions. Here comes into play the Jarzynski equality, as we see in the next section.

\subsection{Jarzynski estimator}

The Jarzynski equality \eref{eq:JE} gives us a recipe to compute the
reversible work, given a suitably-sized collection of irreversible
processes. The work that appears in \eref{eq:JE} is the accumulated work
$W(\Gamma)$ defined in \eref{eq:W(Gamma)}; nonetheless, in some cases it 
happens that the most readily available data for the experimenter is the FEC, therefore the work that is measured is in fact the transferred work $W'(\Gamma)$ of \eref{eq:W'(Gamma)}. In such occasions, the transferred work has been used in the Jarzynski equality, under the assumption that the resulting error is small compared to other sources of experimental uncertainty\cite{Liphardt:2002aa,Collin:2005aa}. 

In this section, we answer the following question: How large an error in the evaluation of $\Delta G_0$ is made if the transferred work $W'(\Gamma)$ is used instead of the accumulated work $W(\Gamma)$?

Let us call $\widetilde{W}$ the Jarzynski estimate of the reversible work $W\sub{rev}$, based on $n$ experiments that produce the set of work measurements $\{W_i\}$:
\be \label{eq:tildeW}
	\beta \widetilde{W}\equiv-\log\sum_{i=1}^n\frac{1}{n}\exp(-\beta W_i) \,.
\ee 
The analogous quantity obtained using the transferred work is
\be \label{eq:W'rev}
	\beta\widetilde{W}'\equiv-\log\sum_{i=1}^n\frac{1}{n}\exp(-\beta W'_i) \,.
\ee
The quantity $\widetilde{W}$ is guaranteed by \eref{eq:JE} to be an estimator of the reversible accumulated work $W\sub{rev}$, whereas $\widetilde{W}'$ is \emph{not} the proper way to compute the reversible transferred work $W'\sub{rev}$ (a bona fide way to estimate $W'\sub{rev}$ is discussed in \rref{Schurr:2003aa}). We now set out to evaluate the difference $\widetilde{W}-\widetilde{W}'$. 

\begin{figure}
   \includegraphics[width=8.6cm]{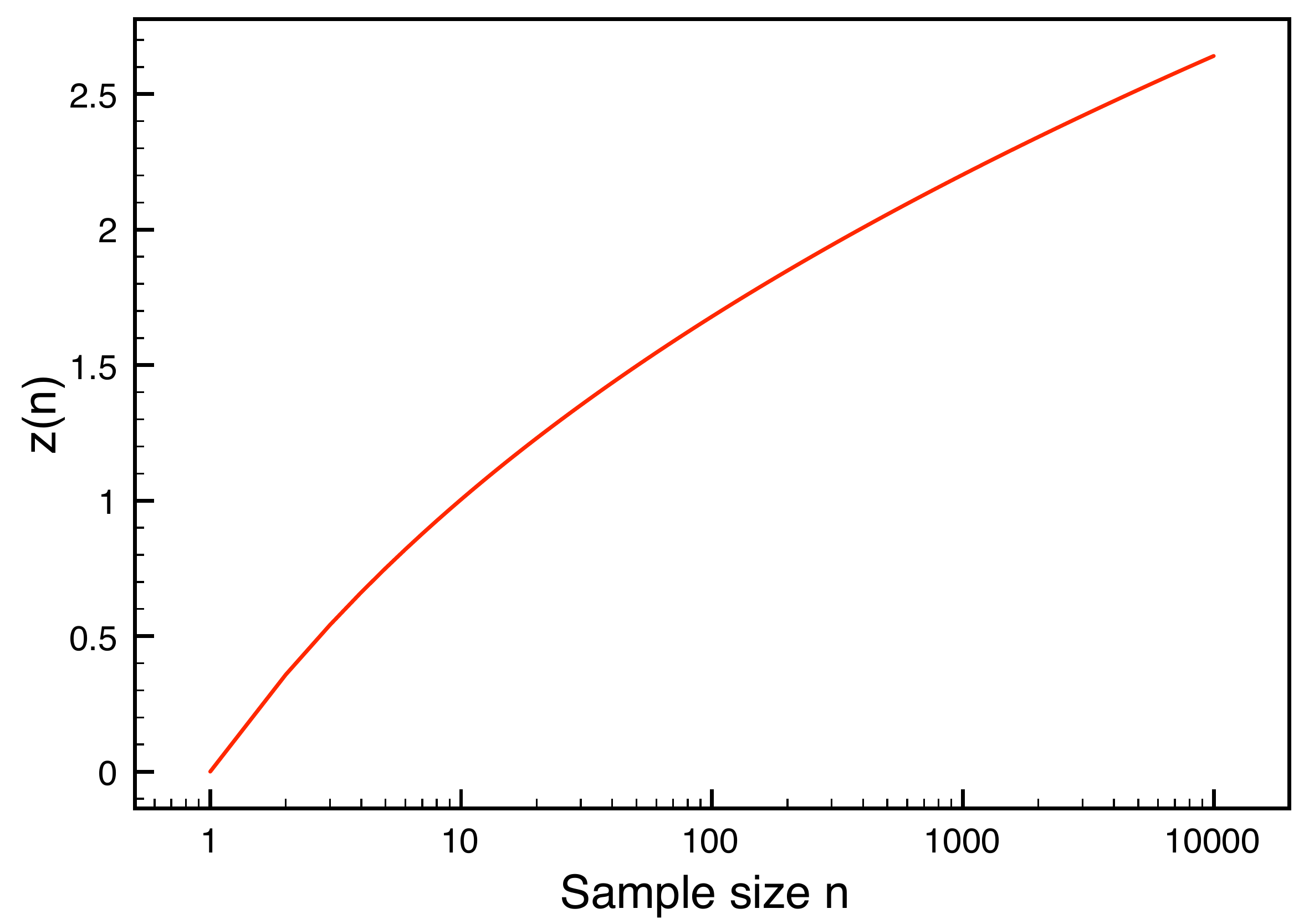} 
    \caption{Dependence on the sample size $n$ of the mode $\omega'$ of
      $W'_{(1)}$ (i.e., the maximum of the distribution for
        $W'_{(1)}$, see \aref{sec:orderstat}). The dimensionless variable $z$ is $(\mu-\omega')/(\sqrt{2}\sigma)$, where $\mu$ and $\sigma$ are the mean and standard deviation of the normally distributed transferred work $w'$. The represented curve is the numerical solution to \eref{eq:Ndep}.} 
   \label{fig:Ndep} 
\end{figure}

To begin with, we sort the set $\{W_i\}$ in ascending order:
\be
	W_{(1)} \leq W_{(2)} \leq W_{(3)} \leq \dots \leq W_{(n)} \,.
\ee 
The key observation is that the sum of exponentials in \eref{eq:tildeW} is dominated by the minimum work trajectory of our sample:
\be
	\beta \widetilde{W}\approx \beta W_{(1)}+\log n \,.
\ee
Repeating the same argument for the set $\{W'_{i}\}$ that collects the measured values of the transferred work, we find
\be
	\widetilde{W}-\widetilde{W}' \approx W_{(1)}-{W'}_{(1)} \,. 
\ee
Note that the trajectory that realizes the minimum of $\{W_i\}$ is generally not the same that gives the minimum of $\{W'_i\}$.

In order to go further in our analytical approximation, we need to specify the distributions of $W$ and $W'$. Based on our experience with both experimental and simulated data, we assume that $W'$ is normally distributed (see \fref{fig:gauss}) with mean $\mu$ and variance $\sigma^2$, while for $W$ we adopt a Gumbel distribution (see \fref{fig:gumbel}) with parameters $a$ and $b$ [which are related to the average and standard deviation of the accumulated work $W$ by means of \eref{eq:aeb} in \aref{sec:orderstat}]. This latter choice is the simplest distribution that exhibits the asymmetry we expect from a nonlinear system\cite{Saha:2007aa} (in the case of linear systems the work distribution is Gaussian\cite{Douarche:2005aa,Douarche:2005ab}). Also, there are theoretical arguments suggesting that  the Gumbel distribution may play a universal role for correlated random variables similar to the one played by the Gaussian distribution for uncorrelated ones\cite{Bertin:2005aa,Bertin:2006aa}.

 We can now estimate the distribution of $W_{(1)}$ and ${W'}_{(1)}$. The details can be found in \aref{sec:orderstat}, here we quote just the final result: the most likely value of $W_{(1)}-{W'}_{(1)}$ is approximately
\be
	a-b\log n-\mu+\sqrt{2}\sigma z(n) \,,
\ee   
where $z(n)$ is the function of the sample size represented in \fref{fig:Ndep}.

What we are really interested in, however, is $\Delta G_0$. If we put $W\sub{rev}=\widetilde{W}$ in \eref{eq:DeltaE} and call $\Delta G_0'$ the result of setting $W'\sub{rev}=\widetilde{W}'$ in \eref{eq:DeltaE'}, we get 
\be
	\Delta G_0-\Delta G_0' \approx a-b\log n-\mu+\sqrt{2}\sigma z(n)-\frac{\langle f\rangle\sub{f}^2-\langle f\rangle\sub{i}^2}{2k\sub{b}} \,.
\ee
A further simplification is possible: taking the average of \eref{eq:W-W'} and using \eref{eq:aeb} we are left with the formula
\be \label{eq:main}
	\Delta G_0-\Delta G_0' \approx \frac{\sqrt{6}}{\pi}(\gamma-\log n)s+\sqrt{2}z(n)s' \,,
\ee
where $s$ and $s'$ are the standard deviations of $\{W_i\}$ and $\{W'_i\}$, respectively, and $\gamma$ is the Euler--Mascheroni constant.

\Eref{eq:main} states that the error in the evaluation of the energy properties of the hairpin due to the substitution of $\{W_i\}$ with $\{W'_i\}$ in the Jarzynski equation depends on three factors: the standard deviations $s$ and $s'$, and the number of experiments $n$. There is a remarkable difference between the roles played by $s$ and $s'$: the standard deviation $s$ of the accumulated work generally depends only on the pulling rate $v$ and the chemical nature of the construct comprising molecule and handles; the standard deviation $s'$ of the transferred work, on the other hand, is also strongly dependent on the bandwidth of the data acquisition system.

The reason is easy to understand: while the area under the FDC [\fref{fig:FDCvsFEC}(a)] practically doesn't change if we smooth out the curve, the area under the FEC [\fref{fig:FDCvsFEC}(b)] is heavily dependent on the fluctuations of the extremal points $x\sub{i}$ and $x\sub{f}$ (see also \fref{fig:FDCvsFEC_exp}). We will have more to say about this point in \sref{sec:exp}.

In the derivation of \eref{eq:main} we have made use of three approximations:
\begin{itemize}
\item we discarded all the contributions to the sum of exponentials in Eqs.\ (\ref{eq:tildeW}) and (\ref{eq:W'rev}) except the one coming from the minimum-work trajectory;
\item we assumed a normal distribution for $\{W'_i\}$;
\item we assumed a Gumbel distribution for $\{W_i\}$.
\end{itemize}
Although each one of them seems reasonable, it is not redundant, before discussing the experimental utility of \eref{eq:main}, to check the final result against a numerical test.

\begin{figure}
   \includegraphics[width=8.6cm]{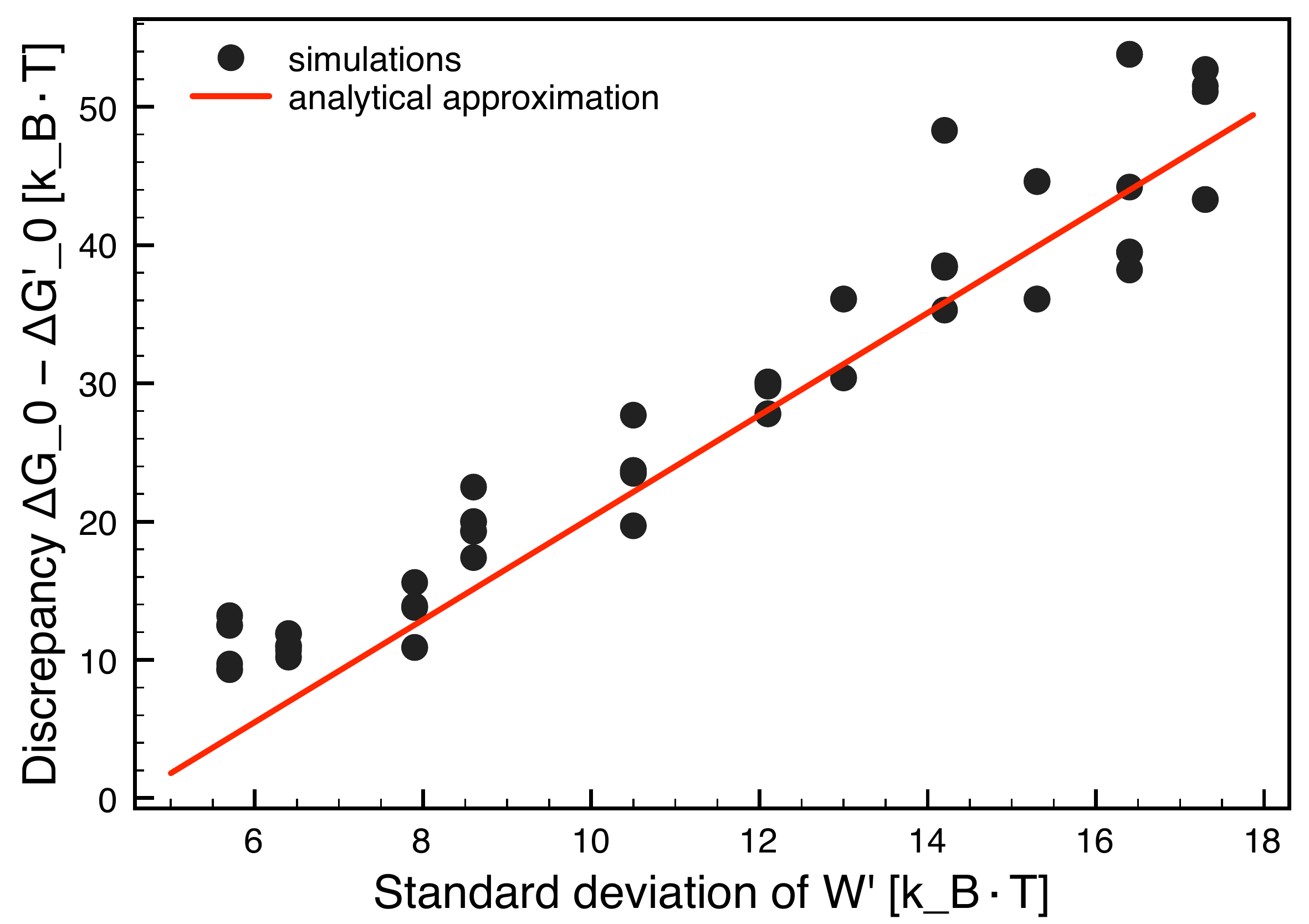} 
    \caption{Numerical test of \eref{eq:main}. The theoretical
      prediction is compared to the results of numerical simulations of
      \eref{eq:LangEq}. In abscissa, $s'$ is the standard deviation of
      the transferred work values $\{W'_i\}$; different values of $s'$
      are obtained by varying the filter applied to the data. In
      ordinate, the error $\Delta G_0-\Delta G_0'$ (in $k\sub{B}T$
        units) on the determination of the free energy of
        formation of the hairpin 
      due to the erroneous use of $W'$ in the Jarzynski estimator. Each
      point represents the result of the analysis of $n=9000$
      trajectories.}
   \label{fig:sgmdep} 
\end{figure}

\subsection{A numerical test}

In order to validate \eref{eq:main}, we have performed a numerical simulation of \eref{eq:LangEq}, generating hundreds of thousands of curves like the two represented in \fref{fig:FDCvsFEC}. The effect of the instrumental bandwidth has been mimicked by applying different filters to the data, so that each point of the FDC or FEC represents actually an average over $m$ consecutive integration steps. In this way we have generated data in a fair range of values of $s'$. The results are illustrated in \fref{fig:sgmdep}. 

The first observation is that the error can be very large: as much as 50
$k\sub{B}T$ in a system where the true $\Delta G_0$ is 57.7 $k\sub{B}T$,
that amounts to a relative error not far from 100\%. Then we observe
that, in spite of the somewhat rough simplifications used in its
derivation, the analytical prediction of \eref{eq:main} fares reasonably
well in the comparison with the simulated data, although there seems to
be a small apparently systematic underestimation of $\Delta G_0-\Delta
G_0'$. Finally, a comment about the range of $s'$: The standard
deviation of $\{W'_i\}$ is a linear function of the amplitude of the
fluctuations of $x$, given by \eref{eq:xfluct}; this fixes an upper
limit to the range of $s'$ that can be explored without changing the
system. 

\begin{figure}   
   \includegraphics[width=8.6cm]{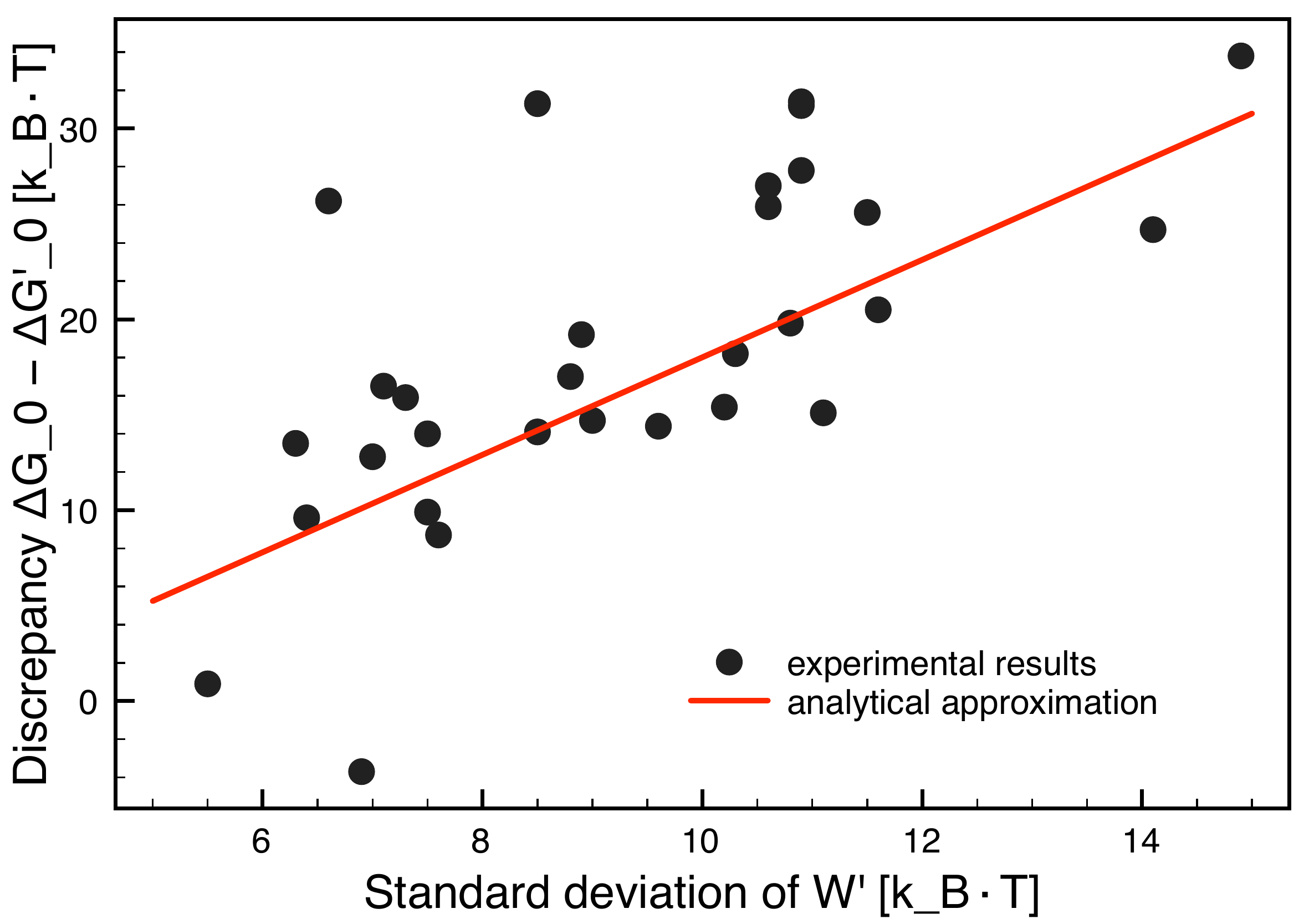}
   \caption{Experimental test of \eref{eq:main}. In abscissa, $s'$ is the standard deviation of the transferred work values $\{W'_i\}$; different values of $s'$ are obtained by varying the stiffness of the trap and the bandwidth. In ordinate, the error $\Delta G_0-\Delta G'_0$ on the determination of the hairpin energy levels due to the erroneous use of $W'$ in the Jarzynski estimator. See \tref{tab:expres2} for further details about the data.}
   \label{fig:expres}
\end{figure}

\section{An experimental test} \label{sec:exp}

This section reports the results of an experimental test of \eref{eq:main}, whose theoretical derivation has been presented in \sref{sec:toy}. The instrument we employed is a dual-beam min\-i\-a\-tur\-ized op\-ti\-cal twee\-zers with fi\-ber-cou\-pled diode lasers (845 nm wavelength) that produce a piezo controlled movable optical trap and measure force using conservation of light momentum\cite{Bustamante:2006aa, Smith:2003aa}. The molecule is a DNA hairpin of sequence 5'-G\-C\-G\-A\-G\-C\-C\-A\-T\-A\-A\-T\-C\-T\-C\-A\-T\-C\-T\-G\-G\-A\-A\-A\-C\-A\-G\-A\-T\-G\-A\-G\-A\-T\-T\-A\-T\-G\-G\-C\-T\-C\-G\-C-3' hybridized to two double-stranded DNA (dsDNA) handles (29 base-pairs long). Pulling experiments were performed at 25~$^\circ$C in a buffer containing Tris H-Cl pH 7.5, 1 M EDTA and 1 M NaCl. The data that we show (see \tref{tab:expres2}) have been measured from 7 specimens in hundreds of stretching-releasing cycles performed at pulling speed of 200 nm/s (equivalent to a loading rate of 13.8 pN/s). The use of DNA hairpins presents several advantages\cite{Woodside:2006aa,Woodside:2006ab,Mossa:2009ab,Mossa:2009ac} over the RNA hairpins that were used in pioneering experiments of this kind\cite{Liphardt:2002aa,Collin:2005aa}.

\begin{table*}
\caption{Experimental results: Comparison between the experimental (also shown in \fref{fig:expres}) and the theoretical (based on \eref{eq:main}) values of $(\Delta G_0-\Delta G'_0)/(k\sub{B}T)$. The datasets labeled ``1 kHz'' and ``20 kHz'' refer to the same experiment, with the standard (low-frequency) and the new (high-frequency) data acquisition system. The stiffness of the trap $k\sub{b}$ is measured in pN/{\textmu}m, while $n$ is the number of trajectories.
\label{tab:expres2}}
\begin{ruledtabular}
\begin{tabular}{cddddddd}
	& & \multicolumn{3}{c}{from unfolding} & \multicolumn{3}{c}{from refolding} \\
	& k\sub{b} & n & \text{exp.} & \text{th.} & n & \text{exp.} & \text{th.} \\
	\hline
	mol1 1 kHz & 79  & 249 & 16.5 & 11.6 & 249 & 8.7 & 12.2 \\
	mol1 20 kHz & 79 & 200 & 31.4 & 20.6 & 202 & 15.1 & 20.2 \\
	mol2 1 kHz & 63 & 473 & 25.6 & 23.3 & 473 & 20.6 & 22.8 \\
	mol2 20 kHz & 63 & 364 & 24.7 & 29.5 & 362 & 33.8 & 31.2 \\
	mol3 1 kHz & 79 & 219 & 15.9 & 17.5 & 218 & 12.8 & 14.7 \\
	mol3 20 kHz & 79 & 169 & 31.2 & 20.1 & 166 & 19.8 & 19.0 \\
	mol4 1 kHz & 64 & 174 & 31.3 & 13.8 & 174 & -3.7 & 9.6 \\
	mol4 20 kHz & 64 & 143 & 18.2 & 17.4 & 138 & 14.4 & 15.8 \\
	mol5 1 kHz & 79 & 635 & 26.2 & 9.1 & 633 & 13.5 & 8.1 \\
	mol5 20 kHz & 79 & 501 & 19.2 & 15.2 & 499 & 17.0 & 14.6 \\
	mol6 1 kHz & 83 & 490 & 14.7 & 15.5 & 492 & 14.1 & 12.6 \\
	mol6 20 kHz & 83 & 386 & 27.0 & 18.6 & 384 & 15.4 & 16.6 \\
	mol7 1 kHz & 77 & 272 & 14.0 & 11.9 & 277 & 9.9 & 11.6 \\
	mol7 20 kHz & 77 & 215 & 27.8 & 20.1 & 219 & 25.9 & 19.1 \\
\end{tabular}
\end{ruledtabular}
\end{table*}

In order to measure the dependence of $\Delta G_0-\Delta G'_0$ on the bandwidth, we employed a fast analog-to-digital converter that makes possible to increase the data acquisition frequency from the standard value of 1 kHz to as much as 100 kHz (20 kHz, however, is larger than the corner frequency of the bead, around 10 kHz, and proved to be enough for this test). The availability of high-frequency data is a good start, but is not enough without a data analysis procedure that carefully preserves the statistical properties of the boundary term [see \eref{eq:W-W'}]. Here are the main steps of the data analysis that we performed:
\begin{enumerate}
\item The stream of data is split into single unfolding or refolding events.
\item Taking advantage of the fact that the elastic response of the short dsDNA handles is with a good approximation Hookean, we fit the FDC folded and unfolded branches with straight lines. 
\item The unavoidable small instrumental drift (which is manifested in the unphysical increasing or decreasing of the measured value of the trap positon $\lambda$) is corrected by shifting the FDC in such a way that the straight line fitting the folded branch crosses $\lambda = 0$ at the same value of the force in any event.
\item The FDCs are integrated between two fixed values $\lambda\sub{i}$ and $\lambda\sub{f}$. These integrations produce two sets of accumulated work values $\{W_i\}$: one for the unfolding and one for the refolding process.
\item Each FEC is integrated between $x\sub{i}(\Gamma)\equiv\lambda\sub{i}-f\sub{i}(\Gamma)/k\sub{b}$ and $x\sub{f}(\Gamma)\equiv\lambda\sub{f}-f\sub{f}(\Gamma)/k\sub{b}$; note that, while $\lambda\sub{i}$ and $\lambda\sub{f}$ are the same for all trajectories, $f\sub{i}$ and $f\sub{f}$ depend on the trajectory $\Gamma$, and so do $x\sub{i}$ and $x\sub{f}$. In this way we obtain two sets of transferred work  values $\{W'_i\}$: again, one for unfolding and one for refolding trajectories.
\item The Jarzynski estimators $\widetilde{W}$ and $\widetilde{W}'$ are computed by means of Eqs.\  (\ref{eq:tildeW}) and (\ref{eq:W'rev}), and then Eqs.\ (\ref{eq:DeltaE}) and (\ref{eq:DeltaE'}) give $\Delta G_0$ and $\Delta G'_0$. 
\end{enumerate}

\begin{figure*}
   \includegraphics[width=8.6cm]{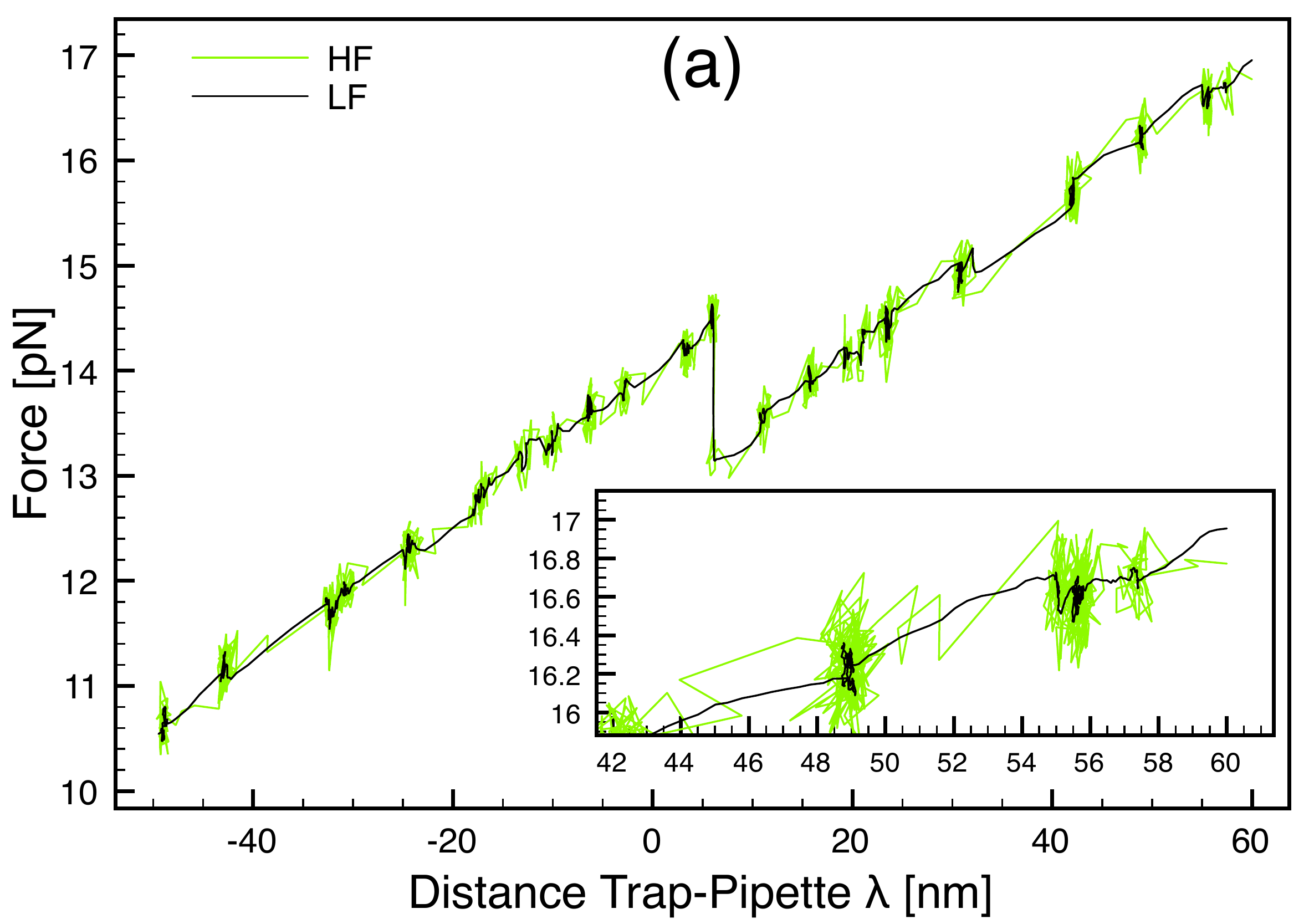} \hspace{0.3cm}
   \includegraphics[width=8.6cm]{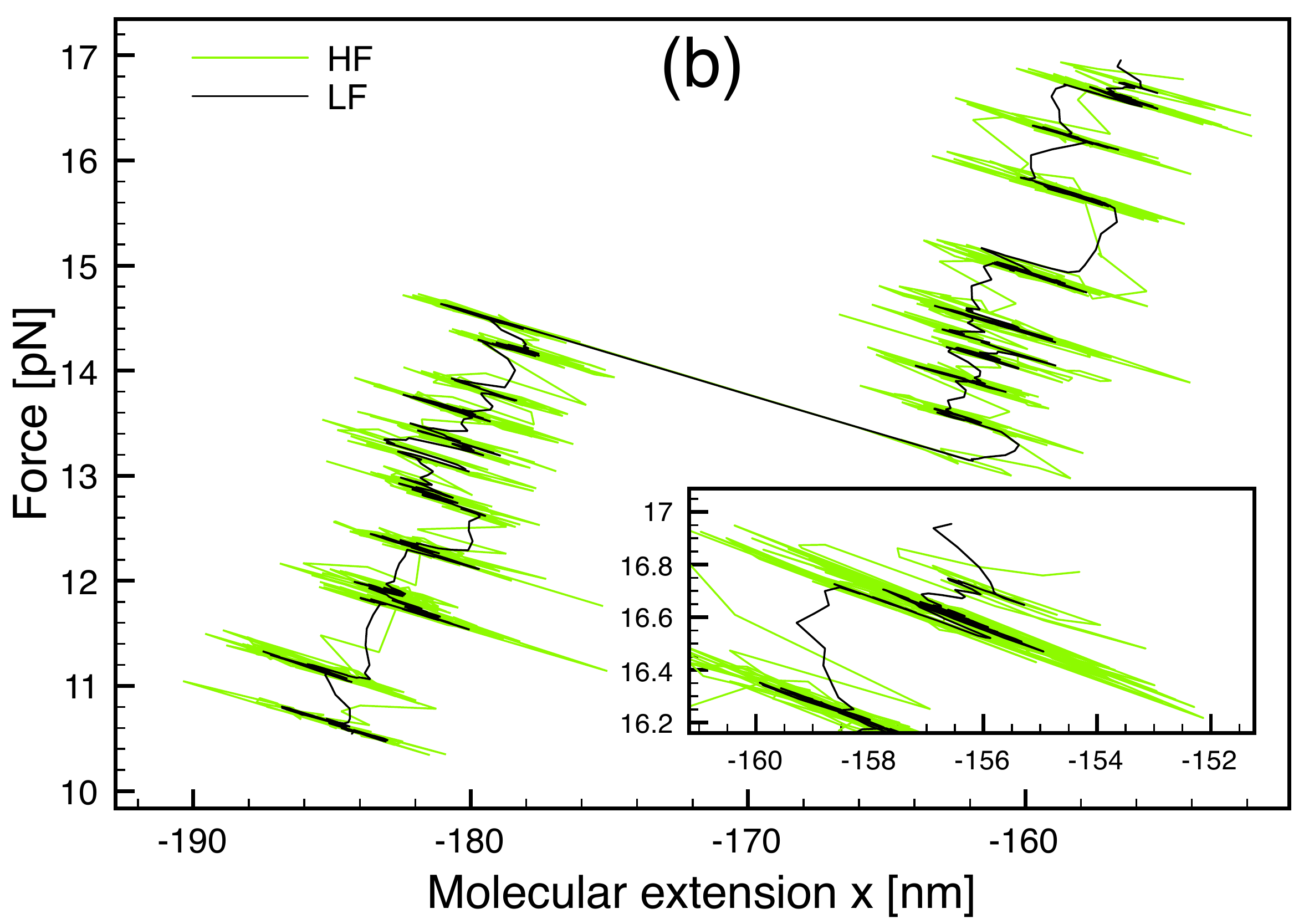}
    \caption{(a) An experimental force-distance curve (FDC) observed with a high-frequency (20 kHz) and a low-frequency (1 kHz) data acquisition system. The area under the curve, which is a measure of the accumulated work $W$, practically doesn't change.  (b) The force-extension curve (FEC) associated to the pulling experiment represented in \fref{fig:FDCvsFEC_exp}(a). The area under the curve, which represents the transferred work $W'$, depends on the frequency of the data acquisition system because of the large fluctuations of the integration extrema. Insets: magnified views of the region around the maximum of the force.} 
   \label{fig:FDCvsFEC_exp} 
\end{figure*}

Table \ref{tab:expres2} shows that \eref{eq:main} is generally quite close to the experimental results, most of the times predicting a discrepancy between $\Delta G_0$ and $\Delta G'_0$ within few $k\sub{B}T$ of the observed value. The occasional large deviations between theory and experiment shouldn't be too surprising in view of the statistical nature of the quantity we are measuring and the approximate derivation of \eref{eq:main}.   

\begin{table*}
\caption{Experimental results. The datasets labeled ``1 kHz'' and ``20 kHz'' refer to the same experiment, with the standard (low-frequency) and the new (high-frequency) data acquisition system. The datasets labeled ``ave $n$'' are obtained from 20 kHz data by averaging over $n$ points.  
\label{tab:expres}}
\begin{ruledtabular}
\begin{tabular}{ldddddd}
	& \multicolumn{2}{c}{from unfolding} & \multicolumn{2}{c}{from refolding} & \multicolumn{2}{c}{bi-directional}\\
	& \Delta G_0 & \Delta G'_0 & \Delta G_0 & \Delta G'_0 & \Delta G_0 & \Delta G'_0 \\
	\hline
	20 kHz & 61.9 & 37.2 & 61.1 & 94.9 & 61.8 & 54.9 \\
	ave 2 & 62.0 & 40.2 & 61.2 & 92.6 & 61.6 & 55.3 \\
	ave 3 & 62.0 & 39.1 & 61.2 & 86.2 & 61.7 & 55.7 \\
	ave 4 & 61.7 & 41.2 & 60.9 & 86.3 & 61.3 & 55.5 \\
	ave 5 & 61.9 & 39.2 & 61.1 & 80.2 & 61.5 & 55.6 \\
	ave 10 & 61.8 & 42.4 & 61.1 & 76.2 & 61.4 & 56.1 \\
	ave 15 & 61.5 & 42.0 & 60.9 & 77.4 & 61.2 & 56.0 \\
	ave 20 & 61.6 & 41.5 & 60.9 & 77.2 & 61.2 & 56.2 \\
\end{tabular} 
\end{ruledtabular}
\end{table*}

The data reported in \tref{tab:expres2} can be graphically represented in analogy with \fref{fig:sgmdep}.
In principle, we expect each dataset  to be represented by a slightly different straight line, as the number of trajectories $n$ varies from a minimum 143 to a maximum 635 (see \tref{tab:expres2}). However, in practice the differences are small enough that all the theoretically expected values are very close to the line that in \fref{fig:expres} is denoted as ``analytical approximation''. 

 \Fref{fig:FDCvsFEC_exp} shows a typical trajectory plotted as FDC and FEC, using 20 kHz and 1 kHz data. It can be immediately appreciated that, while the area under the FDC is insensitive to the sampling frequency, the area under the FEC may display important differences due to the fluctuations of the integration extrema.

\subsection{Bi-directional methods}

\begin{figure}   
   \includegraphics[width=8.6cm]{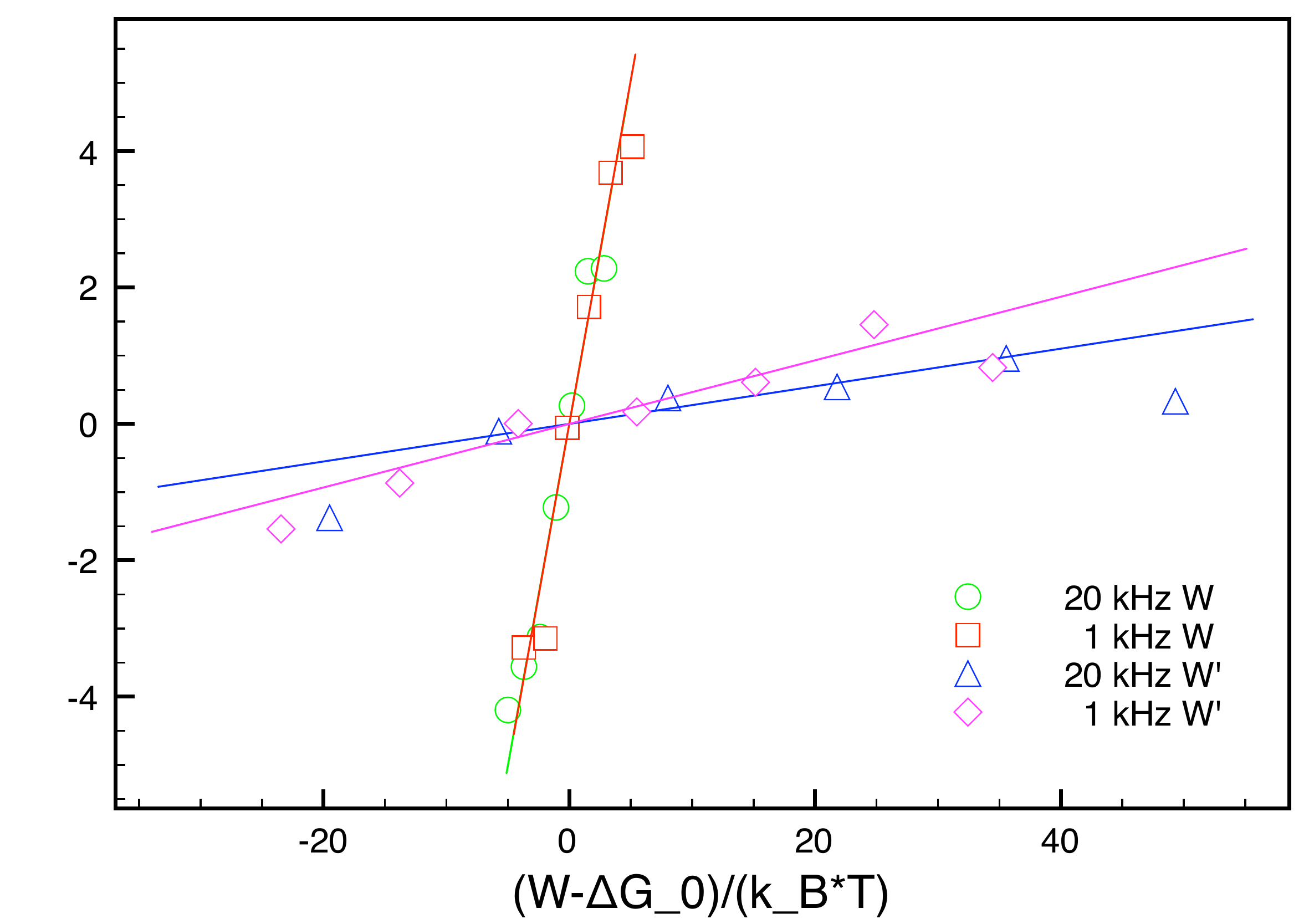}
   \caption{Graph of
       $\log(\phi_{W\sub{FOR}}(w)/\phi_{W\sub{REV}}(-w))$ using high-
       and low-frequency data, accumulated and transferred work. Data
       have been shifted along the horizontal axis to be easily
       compared. Data for the accumulated work (circles and squares)
     fall into a (bandwidth-independent) straight line of
     slope 1.00(8) in quantitative agreement with the
     prediction by the fluctuation relation \eref{eq:Crooks}. However data for the
     transferred work (triangles and rhombs) exhibit bandwidth-dependent
     very small slopes (around 0.03) that exclude the
     validity of an equivalent relation to \eref{eq:Crooks} for
     the transferred work.}
   \label{fig:logplot}
\end{figure}

If the experimental situation makes it possible to implement not only the protocol $\lambda(t)$, but also the time-reversed protocol $\widehat{\lambda}(t)\equiv\lambda(\Delta t-t)$, where $\Delta t\equiv t\sub{f}-t\sub{i}$ is the duration of the experiment, then a more efficient way of estimating free energy differences is to apply a bi-directional method\cite{Shirts:2003aa,Collin:2005aa,Minh:2008aa}, which takes advantage of the knowledge of both a ``forward'' and a ``reverse'' work distributions. Bi-directional methods are based on another fluctuation relation, the Crooks theorem\cite{Crooks:1999aa}
\be \label{eq:Crooks}
	\frac{\phi_{W\sub{FOR}}(w)}{\phi_{W\sub{REV}}(-w)}=\exp\left(\frac{w-\Delta G}{k\sub{B}T}\right) \,,
\ee
where $\phi_{W\sub{FOR}}(w)$ ($\phi_{W\sub{REV}}(w)$) is the probability
density function of the work along the forward (reverse) process. Also
the Crooks theorem, like the Jarzynski equality, is written for the
accumulated work $W$. Writing an analytical approximation of the error
introduced by the erroneous use of the transferred work $W'$, in the
style of what we did in \sref{sec:toy}, looks quite more complicated,
but a direct evidence of the role of the bandwidth is given in
\fref{fig:logplot}, where
$\log[\phi_{W\sub{FOR}}(w)/\phi_{W\sub{REV}}(-w)]$ is plotted as a
function of $(w-\Delta G)/(k\sub{B}T)$ for two values of the bandwith.

The experimental results are summarized in \tref{tab:expres}.
Even if the Crooks theorem is not satisfied, the estimate of $\Delta G_0$ that we get by blindly substituting in \eref{eq:Crooks} the transferred work $W'$ for the accumulated work $W$ is not as bad as the one obtained by using the Jarzynski equality.

\subsection{Role of the data analysis technique} 

The data analysis protocol detailed in \sref{sec:exp} may be the best suited to the task of verifying \eref{eq:main}, but is not feasible if one's experimental setting only provides access to the transferred work $W'$ (and makes it difficult to accurately estimate the stiffness $k\sub{b}$ of the trap). If this is the case, then one either employs a version of the fluctuation theorem written for $W'$ (as in the already cited  \rref{Schurr:2003aa}), or uses $W'$ in \eref{eq:JE}, but takes care of minimizing the error on the determination of $\Delta G_0$, approximately given by \eref{eq:main}. For example, the folded and unfolded branches of the FEC can be smoothed (by application of a filter, by spline-fitting, etc.) until the variance of $\{W'_i\}$ is entirely due to the distribution of the breaking point, in which case the difference between $\Delta G_0$ and $\Delta G'_0$ becomes negligible compared to other sources of experimental error. This is the reason why both Refs.~\onlinecite{Liphardt:2002aa} and \onlinecite{Collin:2005aa} obtained an acceptable experimental test of the Jarzynski equality and the Crooks theorem, respectively, even if erroneously using the transferred work.

\section{Conclusion} \label{sec:end}

The output of a single-molecule pulling experiment can be graphically represented in the form of a force-extension curve, where the externally applied force is compared to the molecular construct end-to-end distance, or a force-distance curve, where the same force is represented against the physical control parameter, the length that can be directly manipulated by the experimenter. The area under the former curve is the work $W'$ transferred to the molecule subsystem, while the latter curve allows the measurement of the accumulated work $W$, the total amount of work expended on the whole system (experimental apparatus included). 

The fluctuation theorems commonly used to compute free energy differences from these out-of-equilibrium processes apply to the work $W$, but not to the work $W'$. In this paper we quantified how large an error is likely to affect the estimate of the free energy at zero force $\Delta G_0$ of the molecule
if $W$ is erroneously replaced with $W'$. We found an analytical
approximated expression [\eref{eq:main}] that emphasizes the role of the
data analysis procedure and of the bandwidth of the data acquisition
system. We confirmed the validity of this approach by both numerical
simulation of a toy model and experiments on a DNA hairpin. This
  work should resolve some issues about the proper way to
  measure work in single-molecule experiments that have generated
  discussion and controversy over the past years.

\acknowledgments
The authors gratefully acknowledge financial support from grants
FIS2007-61433, NAN2004-9348 from Spanish Research Council,  SGR05-00688
from the Catalan Government and RGP55/2008 from Human Frontiers Science Program.
 
\appendix

\section{Thermodynamics of the toy model} \label{sec:therm}

The model defined in \sref{sec:toy} is simple enough to allow the analytical solution of its equilibrium thermodynamics. The partition function of the system is 
\be
	Z(\lambda)=\sum_{\varsigma\in\{0,1\}}\int_{-\infty}^{+\infty}\rmd x\, \exp\left[-\beta H^{(\lambda)}(x,\varsigma)\right] \,,
\ee
where the Hamiltonian is given by \eref{eq:Ham}.
The integration is trivial, so we can immediately write the solution
\be
	Z(\lambda)=Z_0(\lambda)+Z_1(\lambda) \,,
\ee
where
\be
	Z_\varsigma(\lambda)=\sqrt{\frac{2\pi}{\beta k\sub{t}}}\exp\left[-\frac{\beta}{2}k\sub{eff}(\lambda-\ell_\varsigma)^2-\varsigma\beta\Delta G_0\right] \,.
\ee

Given the partition function, we have access to all the thermodynamic properties of the model; the Gibbs free energy, in particular, is defined as 
\be
	G(\lambda)=-k\sub{B}T\ln Z(\lambda) \,,
\ee
and the TFDC is given by
\be \label{eq:TFDC}
	\langle f\rangle(\lambda)=\frac{\partial G(\lambda)}{\partial \lambda}=k\sub{eff}\left[\lambda-P_1(\lambda)\ell_1-P_0(\lambda)\ell_0\right] \,,
\ee
where 
\be
	P_\varsigma(\lambda)=\frac{Z_\varsigma(\lambda)}{Z(\lambda)}
\ee
is the probability of the state $\varsigma$ for a fixed value of $\lambda$. The coexistence value $\lambda\sub{c}$ is characterized by the fact that $P_0(\lambda\sub{c})=P_1(\lambda\sub{c})$, hence
\be	
	\lambda\sub{c}=\frac{\Delta G_0}{k\sub{eff}(\ell_1-\ell_0)}+\frac{\ell_1+\ell_0}{2} \,.
\ee
The corresponding coexistence force is
\be
	f\sub{c}\equiv\langle f\rangle(\lambda\sub{c})=\frac{\Delta G_0}{\ell_1-\ell_0} \,.
\ee
Notice that in the asymptotic region $\lambda\ll \lambda\sub{c}$ the probability of the open state is negligible, so the force goes as $k\sub{eff}(\lambda-\ell_0)$, while in the region $\lambda\gg \lambda\sub{c}$ it is the probability of the closed state that goes to zero, leaving a force dependence of the form $k\sub{eff}(\lambda-\ell_1)$.

From \eref{eq:TFDC} we can easily write down the reversible work 
\be
	W\sub{rev}=\int_{\lambda_i}^{\lambda_f}\langle f\rangle(\lambda)\,\rmd\lambda \,.
\ee
The integration can be done analytically using the fact that
\be
	\int\frac{a\,\rmd x}{a+\rme^{bx}}=x-\frac{1}{b}\ln(a+\rme^{bx}) \,.
\ee
Some tedious algebraic manipulation is required before one can write for the reversible work the following exact formula:
\be \label{eq:Wrev1}
	W\sub{rev}=\Delta G_0+\frac{k\sub{eff}}{2}\left[(\lambda\sub{f}-\ell_1)^2-(\lambda\sub{i}-\ell_0)^2\right]-C \,,
\ee
where $C$ is a correction very small if $\lambda\sub{i}\ll\lambda\sub{c}\ll\lambda\sub{f}$ (that is the most common experimental condition) whose explicit form is
\be
	C=\frac{1}{\beta}\ln\frac{1+\exp[-\beta k\sub{eff}(\ell_1-\ell_0)(\lambda\sub{f}-\lambda\sub{c})]}{1+\exp[-\beta k\sub{eff}(\ell_1-\ell_0)(\lambda\sub{c}-\lambda\sub{i})]} \,.
\ee
In practice, $\lambda\sub{i}\ll\lambda\sub{c}\ll\lambda\sub{f}$ so one can usually forget about $C$ and use \eref{eq:TFDC} to rewrite \eref{eq:Wrev1} as
\be \label{eq:Wrev}
	W\sub{rev}=\Delta G_0+\frac{\langle f\rangle\sub{f}^2-\langle f\rangle\sub{i}^2}{2k\sub{eff}} \,,
\ee
where $\langle f\rangle\sub{i}\equiv\langle f\rangle(\lambda\sub{i})$ and $\langle f\rangle\sub{f}\equiv\langle f\rangle(\lambda\sub{f})$.

The expectation value of the molecular extension is
\be
	\langle x\rangle(\lambda)=\frac{k\sub{b}}{k\sub{t}}\lambda+\frac{k\sub{m}}{k\sub{t}}\left[P_1(\lambda)\ell_1+P_0(\lambda)\ell_0\right] \,.
\ee
This equation can be rephrased into an expression for the TFEC.

Another interesting quantity is the expectation value of $x^2$,
\be
	\langle x^2\rangle(\lambda)=\frac{1}{\beta k\sub{t}}+P_0 x\sub{eq}^2(0)+P_1 x\sub{eq}^2(1) \,,
\ee
from which we easily obtain the variance for the equilibrium fluctuations of $x$
\be \label{eq:xfluct}
	(\delta x)^2(\lambda)=\frac{k\sub{B}T}{k\sub{t}}+P_0(\lambda)P_1(\lambda)\frac{k\sub{m}^2}{k\sub{t}^2}(\ell_1-\ell_0)^2 \,.
\ee
The variance for the equilibrium fluctuations of the force are simply related to those of $x$:
\be 
	(\delta f)^2(\lambda)=k\sub{b}^2(\delta x)^2(\lambda) \,.
\ee

\section{An exercise in order statistics} \label{sec:orderstat}

Let $\{Y_i\}$ be $n$ independent, identically distributed real-valued random variables with cumulative density function (cdf) $\Phi(y)\equiv\Pr(Y_i\leq y)$. The probability density function (pdf) is defined as the derivative of the cdf: $\phi(y)\equiv\Phi'(y)$. The pdf has the property $\phi(y)\rmd y=\Pr(y < Y_i \leq y+\rmd y)$.

The  minimum $Y_{(1)}$ of the set $\{Y_i\}$ is itself a random variable whose distribution can be deduced from the knowledge of $\phi(y)$ and $\Phi(y)$. Indeed, the probability $\Phi_{Y_{(1)}}(y)$ that the minimum is no more than $y$ is equal to the probability of having at least one $Y_i\leq y$. This is given by the binomial distribution as
\be	
	\Phi_{Y_{(1)}}(y)=1-[1-\Phi(y)]^n \,.
\ee 
Differentiating with respect to $y$ we find the corresponding pdf
\be
	\phi_{Y_{(1)}}(y)=n[1-\Phi(y)]^{n-1}\phi(y) \,.
\ee
The simplest way to characterize the most likely value of $Y_{(1)}$ is to consider the mode, that is the point where the pdf has a maximum. This is given by solving with respect to $y$ the following equation:
\be
	[1-\Phi(y)]\phi'(y)=(n-1)\phi^2(y) \,.
\ee  
In the rest of this section, we specialize these general formulas to the two distributions we used to describe the statistical behavior of the accumulated and transferred work.

\subsection{Normal distribution}

\begin{figure}[!t]   
   \includegraphics[width=8.6cm,height=7cm]{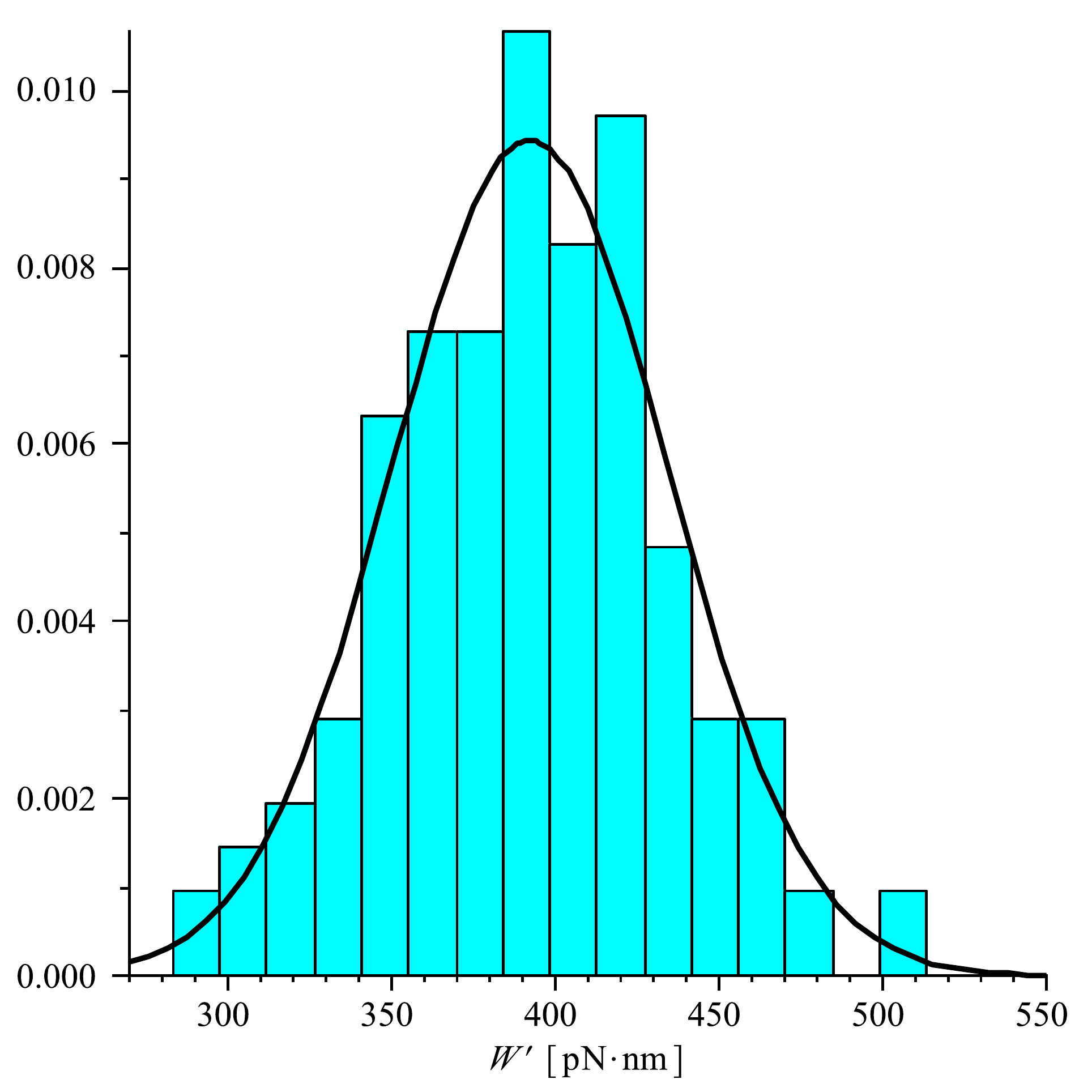}
   \caption{Comparison between the histogram of the transferred work in one of the experiments reported in \tref{tab:expres2} and the normal distribution that better approximates it.}
   \label{fig:gauss}
\end{figure}

A normally distributed variable of mean $\mu$ and variance $\sigma^2$ is described by the cdf 
\be
	\Phi^{\mathrm{(N)}}(y)=\frac{1}{2}+\frac{1}{2}\erf\left(\frac{y-\mu}{\sqrt{2}\sigma}\right) \,,
\ee
from which derives the pdf
\be
	\phi^{\mathrm{(N)}}(y)=\frac{1}{\sigma\sqrt{2\pi}}\exp\left[-\frac{(y-\mu)^2}{2\sigma^2}\right] \,.
\ee
The distribution of the transferred work $W'$ is often well described by a normal distribution (see \fref{fig:gauss}).
It is convenient to define the reduced variable
\be
	z\equiv\frac{\mu-y}{\sqrt{2}\sigma} \,,
\ee
in terms of which we can write the cdf of the minimum $Y_{(1)}$ of a sample of size $n$ 
\be
	\Phi^{\mathrm{(N)}}_{Y_{(1)}}(z)=1-\left[\tfrac{1}{2}+\tfrac{1}{2}\erf (z)\right]^n \,,
\ee
and its pdf
\be \label{eq:phiNxi}
	\phi^{\mathrm{(N)}}_{Y_{(1)}}(z)=\frac{n}{\sigma\sqrt{2\pi}}\exp(-z^2)\left[\tfrac{1}{2}+\tfrac{1}{2}\erf(z)\right]^{n-1} \,.
\ee
The mode of the distribution $\phi^{\mathrm{(N)}}_{Y_{(1)}}(z)$ is the solution to the following transcendental equation:
\be \label{eq:Ndep}
	\sqrt{\pi}z[1+\erf(z)]=(n-1)\exp(-z^2) \,.
\ee 
The numerical solution for $n\leq10\,000$ is plotted in \fref{fig:Ndep}.

\subsection{Gumbel distribution}

\begin{figure}[!th]   
   \includegraphics[width=8.6cm,height=7cm]{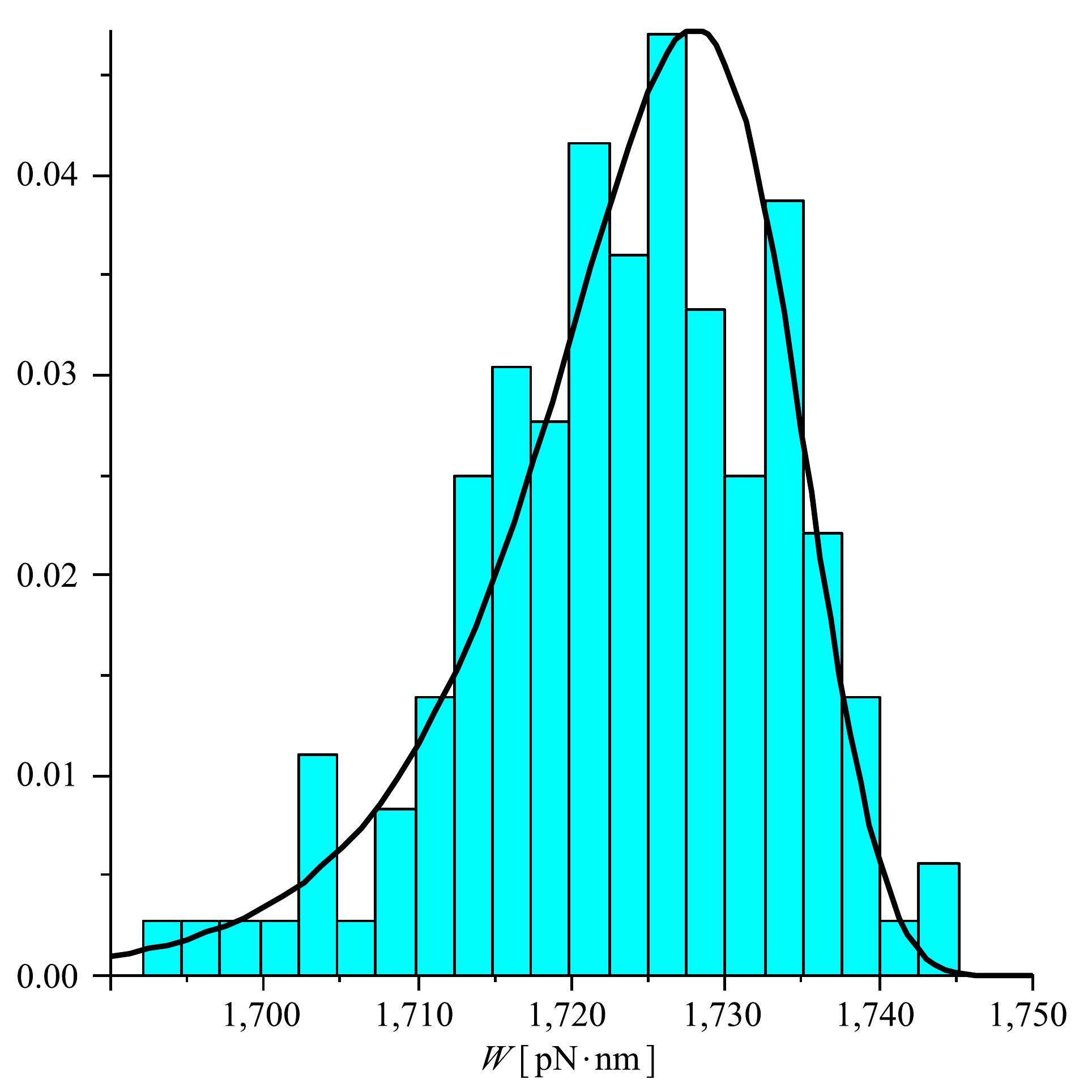}
   \caption{Comparison between the histogram of the accumulated work in one of the experiments reported in \tref{tab:expres2} and the Gumbel distribution that better approximates it.}
   \label{fig:gumbel}
\end{figure}

In both our simulations and experiments, we find that the accumulated work is often adequately represented 
(see \fref{fig:gumbel}) by a random variable obeying the Gumbel distribution
\begin{align}
	\Phi^\mathrm{(G)}(y)&=1-\exp\left[-\exp\left(\frac{y-a}{b}\right)\right] \,, \\
	\phi^\mathrm{(G)}(y)&=\frac{1}{b}\exp\left(\frac{y-a}{b}\right)\exp\left[-\exp\left(\frac{y-a}{b}\right)\right] \,.
\end{align}
The parameters $a$ and $b$ can be quickly estimated from the average $\bar{y}$ and the standard deviation $s$ of the sample $\{Y_i\}$ by means of the formulas
\be \label{eq:aeb}
	b=s\frac{\sqrt{6}}{\pi} \qquad a=\bar{y}+\gamma b \,,
\ee
where $\gamma$ is the Euler--Mascheroni constant 0.5772\dots The minimum value $Y_{(1)}$ over the sample is in this case distributed with pdf
\be
	\phi^\mathrm{(G)}_{Y_{(1)}}(y)=\frac{n}{b}\exp\left(\frac{y-a}{b}\right)\exp\left[-n\exp\left(\frac{y-a}{b}\right)\right] \,.
\ee
The mode of the minimum is therefore given simply by $a-b\log n$.
 
\bibliography{mossa}      

\end{document}